\documentclass[12pt,english]{article}
\usepackage[T1]{fontenc}
\usepackage[latin9]{inputenc}
\usepackage[letterpaper]{geometry}
\usepackage{float}
\usepackage{url}
\usepackage{amsthm}
\usepackage{amsmath}
\usepackage{graphicx}

\makeatletter
\usepackage{amsmath}
\usepackage{amssymb}
\usepackage{url}
\usepackage{listings}
\usepackage{url}

\makeatother

\usepackage{babel}

\begin{document}

\title{Can the Eureqa symbolic regression program, computer algebra and
numerical analysis help each other?}

\author{David R. Stoutemyer%
\thanks{dstout at hawaii dot edu%
}}

\date{February 2012}
\maketitle
\begin{abstract}
The free Eureqa program has recently received extensive press praise.
A representative quote is

{}``There are very clever 'thinking machines' in existence today,
such as Watson, the IBM computer that conquered \emph{Jeopardy}! last
year. But next to Eureqa, Watson is merely a glorified search engine.''

The program is designed to work with noisy experimental data, searching
for then returning a set of result expressions that attempt to optimally
trade off conciseness with accuracy.

However, if the data is generated from a formula for which there exists
more concise equivalent formulas, sometimes some of the candidate
Eureqa expressions are one or more of those more concise equivalents
expressions. If not, perhaps one or more of the returned Eureqa expressions
might be a sufficiently accurate approximation that is more concise
than the given formula. Moreover, when there is no known closed form
expression, the data points can be generated by numerical methods,
enabling Eureqa to find expressions that concisely fit those data
points with sufficient accuracy. In contrast to typical regression
software, the user does not have to explicitly or implicitly provide
a specific expression or class of expressions containing unknown constants
for the software to determine.

Is Eureqa useful enough in these regards to provide an additional
tool for experimental mathematics, computer algebra users and numerical
analysts? Yes, if used carefully. Can computer algebra and numerical
methods help Eureqa? Definitely.
\end{abstract}
\textbf{Keywords}: experimental mathematics, computer algebra, numerical
analysis, simplification, symbolic regression, approximation, searching
algorithms

\medskip{}

\section{Introduction\label{sec:Introduction}}

Eureqa is a highly praised symbolic regression program described by
Schmidt and Lipson \cite{SchmidtAndLipson}, freely down-loadable
from \cite{Schmidt}, where there is a bibliography of its use in
articles, press citations, a blog and a discussion group. See \cite{KozaExampleOfGeneticProgramming}
for quick insight into the underlying Darwinian method for symbolic
regression. Eureqa was designed to work from noisy experimental data.
However in the first few weeks of using this tool I have already found
that also:\vspace{-5pt}

\begin{enumerate}
\item Eureqa can sometimes do a better job than existing computer algebra
systems of exact simplification if an exact result isn't too complicated;\vspace{-5pt}

\item Eureqa can often discover simple expressions that \textsl{approximate}
more complicated expressions or fit a set of numerical results well;\vspace{-5pt}

\item Even when the fit isn't as accurate as desired, the \textsl{form}
of a returned expression often suggests a class of forms to try for
focused routine regressions, interpolations or series expansions giving
higher accuracy; and\vspace{-5pt}

\item Eureqa could do an even better job if it supplemented its search with
exploitation of more computer algebra and standard numerical methods,
including classic regression;\vspace{-5pt}

\item There are important caveats about how to use Eureqa effectively.\vspace{-5pt}

\end{enumerate}
This article has examples that illustrate these findings, using Eureqa
0.93.1 beta.

Regarding computing times reported herein, the computer is a 1.60GHz
Intel Core 2 Duo U9600 CPU with 3 gigabytes of RAM.

\section{Exact simplification and transformation}

\subsection{A trigonometric simplification example\label{sub:A-trigonometric-simplificationEg}}

Here is a Maple 15 assignment of an input trigonometric expression
to a dependent variable $y$:\vspace{-5pt}
\begin{multline}
y\,:=\,\cos(x)^{3}\sin(x)+\dfrac{\cos(x)^{3}\sin(x)}{2}+2\cos(x)^{3}\cos(2x)\sin(x)+\dfrac{\cos(x)^{3}\cos(4x)\sin(x)}{2}-\\
\frac{3}{2}\cos(x)\sin(x)^{3}-2\cos(x)\cos(2x)\sin(x)^{3}-\dfrac{\cos(x)\cos(4x)\sin(x)^{3}}{2}.\quad\label{eq:MessyTrigExpression}\end{multline}
The default simplification merely combines the first two terms, which
are similar.

However, then $\mathrm{simplify}(y)$ required only 0.03 seconds to
return the much simpler\vspace{-5pt}
\begin{equation}
4\sin(x)\cos(x)^{5}\left(2\cos(x)^{2}-1\right).\label{eq:MapleSimplifyTrig}\end{equation}
An equivalent expression produced by the \textsl{Mathematica} $\mathrm{FullSimplify}[\ldots]$
function in only 0.08 seconds is also much simpler:%
\footnote{\textsl{Mathematica} output in this article is shown as produced by
ending the input with {}``// TraditionalForm''%
}\vspace{-5pt}
\begin{equation}
2\left(\sin(3x)-\sin(x)\right)\cos^{5}(x)\,.\label{eq:FullSimplifyTrig}\end{equation}
But the equivalent even simpler form discovered by Eureqa is\vspace{-5pt}
\begin{equation}
y\;=\;\cos(x)^{4}\sin(4x)\,.\label{eq:NiceExactTrigResult}\end{equation}
\textsl{Caveat}: The algorithm uses a random number generator with
no user control over its seeding, which is done perhaps by the clock.
Therefore sequences are not currently repeatable and the computing
times to obtain expression (\ref{eq:NiceExactTrigResult}) varied
dramatically, from 3 seconds to several minutes. It conceivably could
require more time than you would ever be willing to invest. However,
although even 3 seconds is much longer than the computer algebra times,
it is certainly worth a few minutes wait to obtain such a nice result,
and such experiments can be done while away from the computer, such
as eating, sleeping or playing cell-phone games. There is also an
option to use parallel cloud computing, which reduces the mean and
variance of the elapsed time necessary to obtain a satisfactory result.

\begin{figure}[H]
\noindent \centering{}\caption{Eureqa \textsf{Set Target }tab for exact trigonometric simplification
example}
\label{Flo:SetTargetTab}\includegraphics[bb=0bp 0bp 617bp 608bp]{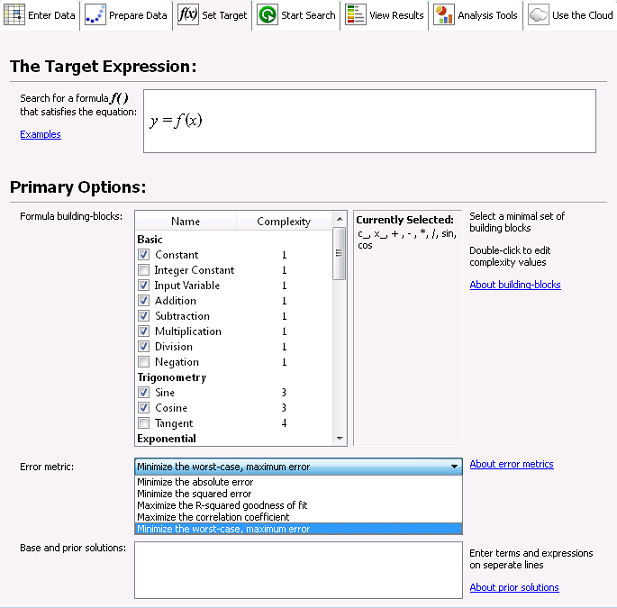}
\end{figure}

Here is how I used Eureqa to obtain this delightfully simple exact
result (\ref{eq:NiceExactTrigResult}):\vspace{-5pt}

\begin{enumerate}
\item First I plotted expression (\ref{eq:MessyTrigExpression}) in \textsl{Mathematica},
revealing that it is antisymmetric and that its fundamental period
appeared to be $\pi$, with higher frequency components appearing
to have a minimum period of $\pi/4$. This was confirmed by $\mathrm{TrigReduce}[y]$,
which returned the equivalent expression\vspace{-5pt}
\begin{equation}
\frac{1}{16}\left(\sin\left(2x\right)+6\sin(4x)+4\sin\left(6x\right)+\sin(8x)\right)\,.\label{eq:TrigReduceOfInput}\end{equation}
\vspace{-5pt}

\item I decided to use evenly-spaced values of $x$ because expression (\ref{eq:MessyTrigExpression})
is periodic, bounded and $C^{\infty}$ for all real $x$. I guessed
that it might help Eureqa discover and exploit the antisymmetry if
I used sample points symmetric about $x=0$. I also guessed that it
might help Eureqa discover the periodicities if I used exactly two
fundamental periods. I guessed that using 16 samples within the minimum
period $\pi/4$ would be sufficient to resolve it quite well, then
I doubled that because Eureqa uses some points for fitting and others
for error assessment. This works out to 128 intervals of width $\pi/64$
from $-\pi$ through $\pi$. I then created a table of 17-digit $x$
$y$ floating-point pairs and exported it to a comma-separated-values
file by entering\vspace{-5pt}
\[
\mathrm{Export}\left[\mathtt{"}\mathrm{trigExample.csv}\mathtt{"}\mathrm{,\, Table\,}[\mathrm{N\,}[\{x,\ldots\},17],\{x,-\pi,\,\pi,\frac{\pi}{128}\}]\right]\]
where the ellipsis was expression (\ref{eq:MessyTrigExpression}).%
\footnote{The reason for requesting 17-digits was that I wanted \textsl{Mathematica}
to use its adaptive significance arithmetic to give me \textsl{results}
estimated accurate to 17 digits despite any catastrophic cancellations,
and I wanted Eureqa to receive a 17th significant digit to help it
round the sequences of input digit characters to the closest representable
IEEE double values, which is used by Eureqa. It might appear that
I am being compulsive here. However, the Eureqa result could contain
floating point constants that are approximations to constant expressions
such as $\pi/6$ or $\sqrt{3}/2$. Whenever software contains a floating
point number that I suspect might closely approximate such an exact
constant, I submit the floating point number to the amazing free Internet
Inverse Symbolic Calculator \cite{InverseSymbolicCalculator-1}. The
complexity of the constants that it can identify increases with the
precision of the floating point number, and 16 digits of precision
is near the minimum necessary to identify linear functions of $\pi$,
$e$, and square roots of integers with rational coefficients. Maple
contains an earlier version of Inverse Symbolic Calculator named $\mathrm{identify}(\ldots)$.

The $y$ values could be generated from the $x$ values \textsl{within}
the Eureqa \textsf{Enter Data} tab by entering the formula {}``$=\ldots$'',
where the ellipsis is the right side of formula (\ref{eq:MessyTrigExpression}).
However, Eureqa rounds such formula-generated values to 15 significant
digits. The $x$ and $y$ values can also be generated in Excel, then
copied and pasted into Eureqa. However, Excel tends to round values
to 10 significant digits. Such rounding might be welcome for aesthetic
reasons with low-accuracy experimental data, but it is an unnecessary
loss of accuracy for experimental mathematics.%
}\vspace{-5pt}

\item I then launched Eureqa, opened its spreadsheet-like \textsf{Enter
Data} tab, then replaced the default data there with mine by importing
file trigExample.csv. In the row labeled \textsf{var} I then entered
$x$ in column A and $y$ in column B.\vspace{-5pt}

\item The \textsf{Prepare Data} tab then showed superimposed plots of the
$x$ and $y$ data values and offered preprocessing options that aren't
relevant for this very accurate data.\vspace{-5pt}

\item Figure \ref{Flo:SetTargetTab} shows the \textsf{Set Target} tab:\vspace{-5pt}

\begin{enumerate}
\item The pane labeled \textsf{The Target Expression} suggests the fitting
model $y=f(x)$, which is exactly what I want, so I didn't change
it.\vspace{-3pt}

\item The pane labeled \textsf{Primary Options} has check boxes for the
desired \textsf{Formula building blocks} used in composing candidate
$f(x)$ expressions. The default checked ones are sufficient for the
sort of concise equivalent to expression (\ref{eq:MessyTrigExpression})
that I am seeking. Therefore I didn't check any of the other offered
functions and operators -- not all of which are shown on this screen
shot.%
\footnote{I could have saved search time by un-checking \textsf{Division}, because
the floating point coefficients make a denominator unnecessary for
this class of expressions. I could also have saved search time and
perhaps obtained a more accurate result by checking the \textsf{Integer
Constant} box, because integer coefficients are quite likely for exact
equivalent expressions, making it worth having Eureqa try rounding
to see if that improves the accuracy. However, I decided to accept
the default checked boxes to see if Eureqa could find a good exact
equivalent without any more help from me. Other building blocks are
described at \cite{EureqaBuildingBlocks}.%
} That dialog pane also shows the corresponding building-block \textsf{Complexity}
measures, which can be altered by the user. The complexity measure
of an expression is the total of the complexities of its parts.\vspace{-3pt}

\item The drop-down menu labeled \textsf{Error metric} offers different
built-in error measures for Eureqa to try optimizing. A bound is usually
more reassuring than the alternatives, so I chose \textsf{Minimize
the worst-case maximum error}. The documentation suggests that \textsf{Maximize
the R-squared goodness of fit} or \textsf{Maximize the correlation
coefficient} is more scale and offset invariant, which are desirable
properties too. However, the data is already well scaled with means
of 0.\vspace{-3pt}

\item For the covered \textsf{Data Splitting} drop-down menu the default
alternative is to designate a certain percentage of the data points
for fitting the data (\textsl{training}) and a certain percentage
for assessing the error measure (\textsl{validation}). The individual
assignments to these categories are done randomly, with some overlap
if there aren't many data points. For the almost exact data in this
article I would prefer that alternate points be assigned to alternate
categories, with the end points being used for training. However,
none of the alternatives offered this so I chose the default.%
\footnote{I have since learned that with such precise data another promising
alternative would be to use all of the points for both training and
validation, which is a current option.%
}\vspace{-5pt}

\end{enumerate}
\item I then pressed the \textsf{Run} button on the \textsf{Start Search}
tab and watched the temporal progress of the search. The plot in Figure
\ref{Flo:StartSearchTab} dynamically zoomed out to show a log-log
plot of the lower left envelope of the most accurate candidate obtained
so far as a function of computing time. The \textsf{Project Log} dynamically
showed successive candidates that are better than any so far on the
basis of either complexity or accuracy. The entertainment of watching
the evolution of these panes is very appealing. It is similar to rooting
for the home team at a sporting event. Notice that:\vspace{-5pt}

\begin{enumerate}
\item Integer powers of sub-expressions are represented in the Project Log
as repeated multiplications, and their complexity is measured that
way. For example, several results are displayed as\vspace{-5pt}
\[
y=\cos(x)*\cos(x)*\cos(x)*\cos(x)*\sin(4\!*\! x)\]
with a complexity measure of 26.%
\footnote{The complexity measure seems less than ideal -- the \textsl{ultimately}
reported formula in the \textsf{View Results} tab of Figure \ref{Flo:ViewResultsTab}
is $\cos(x)^{4}\sin(4x)$. This is more concise and requires only
\textsl{one} cosine and \textsl{two} multiplies to compute the $\cos(x)*\cos(x)*\cos(x)*\cos(x)$
factor if compiled by any decent compiler. %
}\vspace{-3pt}

\item This result appears first with a reported \textsf{Fit} of 0.00111,
followed by the same expression with monotonically better fits up
through $3.81\times10^{-10}$ at 39 seconds.%
\footnote{The fact that the same displayed formula was associated with such
dramatically different error measures is because Eureqa always round
its \textsl{displayed} coefficients to about four significant digits,
and in this case it enabled display of an exact result that it was
merely converging on. If I had checked the \textsf{Integer Constant}
building block, then Eureqa probably would have rounded the coefficients
to integers. However, $y(0.0)=0.0$ and the data varies between about
$\pm3.8$, so a maximum residual of $3.81\times10^{-10}$ is enough
to convince most people that the \textsl{displayed} result is exact.
This can be proved by entering\[
\mathrm{TrigReduce}[\ldots-\mathrm{Cos}[x]^{4}\,\mathrm{Sin}[4x]]\]
with the ellipsis being expression (\ref{eq:MessyTrigExpression}).
This input returns 0, proving equivalence.%
} I terminated the search at 8 minutes, during which Eureqa tried different
formulas that didn't fit as well no matter how complicated they were.\vspace{-5pt}

\end{enumerate}
\item Figure \ref{Flo:ViewResultsTab} shows the results on the \textsf{View
Results }tab: \vspace{-5pt}

\begin{enumerate}
\item In the pane labeled \textsf{Best Solutions of Different Sizes}, the
first column lists the complexity measure, the second column lists
an error measure, and the third column lists the corresponding candidate
equation. If you click on a row, the pane in the lower left corner
gives more detail, with various rounded error measures including the
\textsf{Primary Objective} that I chose, which was maximum absolute
error.\vspace{-5pt}

\item Eureqa does some automatic expression simplification: Whenever a \textsl{mutation}
produces a new candidate expression or a \textsl{crossover} produces
two new expressions containing mixtures of the two parents' sub-expressions,
a few transformations are applied to make the expression more nearly
canonical and to avoid having a misleadingly high complexity on account
of uncollected terms, uncombined numeric sub-expressions, etc. This
minimal computer algebra is also used to make the displayed results
in the \textsf{Best solutions of different sizes} pane more attractive.
These transformation include:\vspace{-3pt}

\begin{itemize}
\item Integer powers and products of sums are expanded to make them more
nearly canonical.
\item Factors and terms are sorted, then similar factors and terms are collected.
\item Numeric sub-expressions are reduced to a single number.
\item A few rules such as $\mathrm{abs}(\mathrm{abs}(u))\rightarrow\mathrm{abs}(u)$
and $1*u\rightarrow u$ are applied.
\end{itemize}
However, you will often see a sub-expression such as $1\cos(1x)$.
This is because the 1s are a result of rounding non-integer coefficients
to the typically-displayed 3 or 4 significant digits. Nonetheless
I noticed definite unexploited opportunities such as Eureqa unnecessarily
trying the sub-expression $\sin(x+6.283)$. Angle reduction and exploitation
of odd or even symmetry could preclude the need to try candidates
having constant terms outside a much smaller interval. Perhaps rather
than gradually implementing more custom computer algebra, Eureqa could
embed one of the many free computer algebra systems compared at \cite{ComparisonOfCAS},
many of which are conveniently down-loadable from \cite{sourceforgeCAS}.

\item The pane in the lower right corner plots an error measure as a function
of complexity for the 12 reported optimal candidates, with the currently
highlighted candidate as a larger red dot.\vspace{-5pt}

\item The pane in the upper right corner shows a plot of the highlighted
expression superimposed on the validation points in dark blue and
the training points in lighter blue. The drop-down menu above it offers
the option of instead plotting the residuals at all of the data points,
which gives a much better idea of the fitting errors as a function
of $x$. If these residuals had revealed a non-negligible recognizable
pattern such as being approximately of the form $\sin(16x)$, then
I would have tried another search with the model $y=f(\cos(x),\sin(4x),\sin(16x))$.
I would iterate this process until there was no further improvement
or the residuals revealed no non-negligible recognizable pattern.\vspace{-5pt}

\end{enumerate}
\end{enumerate}
Instead of searching for the most concise equivalent, one can also
search for equivalent expressions of a \textsl{particular form}. For
example, we could request that expression (\ref{eq:MessyTrigExpression})
be transformed into a function without cosines by un-checking that
building block. As another example, given an expression $z=\ldots$
that depends on $x$ and $y$, perhaps it is desired to transform
it to an expression depending only on $x$ times an expression depending
only on $y$. This can be requested by making the target expression
be $z=f_{1}(x)*f_{2}(y)$. (Nonnegative integer suffixes on variable
and function names are pretty-printed as subscripts.)

\subsection{Some other exact simplification examples}

Using {}``$\mathrm{Assuming[\ldots,\, FullSimplify}[\ldots]]$''
in \textsl{Mathematica} and {}``$\mathrm{simplify}(\ldots),\,\mathrm{assuming}\,\ldots$''
in Maple, I could not make them accomplish the following quick successful
Eureqa simplifications, for which I selected the default building
blocks,\textsf{ Integer Constants}, and other functions or operators
that occur in the specific given expressions:\vspace{-5pt}
\begin{eqnarray}
\left\lfloor \sqrt{\left\lfloor x\right\rfloor }\right\rfloor +\left\lfloor \sqrt{x}\right\rfloor \;|\; x\geq0 & \rightarrow & 2\left\lfloor \sqrt{x}\right\rfloor .\label{eq:SqrtFloor}\end{eqnarray}
\vspace{-5pt}
\begin{equation}
q^{2}\!+\!\frac{2^{1/3}\!\left(\!\sqrt{81q^{2}\!-\!12}\!-\!9q\right)^{2/3}\!+\!24^{1/3}}{6^{2/3}\left(\sqrt{81q^{2}-12}-9q\right)^{1/3}}-\frac{2\cos\!\left(\!\frac{1}{3}\mathrm{acos\!}\left(\!-\frac{3\sqrt{3}q}{2}\right)\right)}{\sqrt{3}}\;\mid\;\left|q\right|\!<\!\frac{2}{3\sqrt{3}}\rightarrow q^{2}.\label{eq:Cubic2Ways}\end{equation}
\vspace{-5pt}
\begin{eqnarray}
\max(x\!-\! y,\,0)-\max(y\!-\! x,\,0) & \rightarrow & x-y.\label{eq:MaxExample}\end{eqnarray}

\begin{itemize}
\item Example (\ref{eq:SqrtFloor}) comes from \cite{Concrete Mathematics}.\vspace{-5pt}

\item I tabulated only the \textsl{real} part for example (\ref{eq:Cubic2Ways})
because rounding errors for the principal branch of the fractional
powers of the negative quantity $\sqrt{81q^{2}-12}-9q$ generated
some relatively very small magnitude imaginary parts.\vspace{-5pt}

\item To generate $17^{2}\rightarrow289$ rows of data for example (\ref{eq:MaxExample}),
I used\vspace{-5pt}
\begin{multline*}
\mathrm{Apply}[\mathrm{Join},\mathrm{\! Table}[\mathrm{N}[\{x,y,\mathrm{Max}[x\!-\!\frac{2}{3\sqrt{3}}y,0]\!-\!\mathrm{Max}[y\!-\! x,0]\}\!,17],\\
\{x,-1,1,\frac{1}{8}\}\!,\{y,-1,1,\frac{1}{8}\}].\end{multline*}

\end{itemize}
The examples so far illustrate that Eureqa can sometimes determine
a simpler exact result than a computer algebra system. However, instead
of choosing random examples or ones from the literature, I constructed
these very simple results, then obfuscated them in ways that I thought
would be difficult for common transformations to reverse -- particularly
if applied to the entire expression rather than well chosen pieces.
Despite this it was not easy to find examples for which neither $\mathrm{FullSimplify}[\ldots]$
nor $\mathrm{simplify}(\ldots)$ could determine the very simple equivalent.
Thus, Eureqa should be regarded as an occasionally beneficial supplement
to these functions rather than a replacement. Also, effective use
of Eureqa requires good judgment in:\vspace{-5pt}

\begin{itemize}
\item the interval spanned by the sample points, their number and spacing;\vspace{-5pt}

\item the form selected for the target expression;\vspace{-5pt}

\item the Building blocks that are selected together with their complexities,
and\vspace{-5pt}

\item perhaps the error measure that is selected.\vspace{-5pt}

\end{itemize}
Effective use also probably depends on experience with Eureqa. Therefore
the benefits that I have discovered should be regarded as a lower
bound because of my novice status.

\subsection{How to reduce the curse of dimensionality}

As illustrated by example (\ref{eq:MaxExample}), data for representing
all combinations of $n$ different values for each of $m$ different
independent variables requires $n^{m}$ rows, and we must have $n\geq2$
for Eureqa to discern any non-constant dependence on each variable.
Thus for a given $n$, this exponential growth in data with the number
of independent variables can greatly increases the time required for
Eureqa to find a good fit. Yielding to the consequent temptation to
save time by reducing $n$ as $m$ increases tends to reduce the precision
of the results. Fortunately, there are several complementary techniques
that sometimes reduce the effective dimensionality:\vspace{-5pt}

\begin{enumerate}
\item \textsl{Dimensional analysis} can sometimes reduce the number of independent
variables, and \cite{StoutemyerDimensionalAnalysis} describes a Maxima
program that does this.\vspace{-5pt}

\item If you can partition an expanded form of your expression into two
or more sums having disjoint variable sets, then you can independently
fit each of those sub-sums.\vspace{-5pt}

\item If you can partition a factored form of your expression into two or
more products having disjoint variable sets, then you can independently
fit each of those sub-products.\vspace{-5pt}

\item If in every term of a given sum the exponents of some subset of the
variables sum to the same homogeneity exponent $k$, then substitute
1 for one of those variables $v$, fit this isomorphic problem, then
multiply every resulting term by the individual power of $v$ necessary
to restore homogeneity exponent $k$\vspace{-5pt}

\item Perhaps some other change of variables such as a matrix rotation can
totally or partially decouple the independent variables.
\item Unlike many experimental situations, for experimental mathematics
we usually have complete freedom to choose the data values of the
independent variables. In both univariate and multivariate problems
there can be \textsl{sweet spots} that deliver more accuracy for a
given number of samples than does a brute force Cartesian product
of uniformly spaced points. These special values are often related
to the zeros or extrema of orthogonal polynomials. For example, see
the the multidimensional integral formulas in Chapter 25 of \cite{AbramowitzAndSegun}.
\end{enumerate}

\section{Approximate symbolic results\label{sec:Approximate-symbolic-results}}

\begin{flushright}
\vspace{-5pt}

\par\end{flushright}

Ideally a floating point result is either exact or the closest representable
number to the exact result, with ties broken in favor of having the
last significand bit be 0. Well designed floating-point \textsl{arithmetic}
comes very close to this ideal: If the exact result is representable
within the thresholds for overflow and for gradual underflow, then
the result is the exact result for inputs that differ from the actual
inputs by factors between $1\pm\varepsilon_{m}$, where \textsl{machine
epsilon} $\varepsilon_{m}$ is about $1.11\times10^{-16}$ for IEEE
double, which corresponds to about 16 significant digits. A relative
error bound of $1\pm\varepsilon_{m}$ is the \textsl{gold standard}.
There are similar expectations for operations such as exponentiation
and functions such as sinusoids or logarithms that are commonly built
into compilers -- but allowing a few times $\varepsilon_{m}$, which
I call the \textsl{silver standard}. It is a matter of professional
pride among numerical analysts designing \textsl{general purpose}
mathematical software to strive for the silver standard for all functions
provided with the software, such as Bessel functions, etc. I was curious
to know if Eureqa can help numerical analysis by discovering concise
approximate expressions that meet the silver standard.

\subsection{An antiderivative example\label{sub:An-antiderivative-example}}

Suppose that I want to implement an IEEE double version of the Dogbert
$W$ function defined by\vspace{-5pt}
\begin{eqnarray}
W(x) & := & \intop_{0}^{x}\dfrac{dt}{\sqrt{1-t^{2}+\frac{2}{\pi}\cos\left(\frac{\pi t}{2}\right)}}.\label{eq:DesiredAntiderivative}\end{eqnarray}
No computer algebra systems that I tried can determine a closed form
for $W(x)$.

\textsl{Plan B}: Do a gold-standard numeric integration for a sequence
of $x$ values in the interval -1$\leq x\leq1$, then use Eureqa to
discover a silver-standard expression that fits those values well.
Here is an abridged account of my attempt to do this.\vspace{-5pt}

\begin{enumerate}
\item I entered the \textsl{Mathematica} input\vspace{-5pt}
\begin{multline}
\mathrm{xWPairs}\;=\;\mathrm{Table}\,[\{\mathrm{N}\,[x,17],\,\mathrm{NIntegrate}\,[\intop_{0}^{x}\dfrac{1}{\sqrt{(1-t)(1+t)+\frac{2}{\pi}\cos\left(\frac{\pi t}{2}\right)}},\;\\
\{t,0,x\}\mathrm{,PrecisionGoal}\rightarrow17,\,\mathrm{WorkingPrecision}\rightarrow25],\{x,-1,1,1/64\}.\enskip\label{eq:xyPairs}\end{multline}
Then I inspected the 127 number pairs to make sure there were no imaginary
parts, infinities or undefined values.%
\footnote{\begin{itemize}
\item The data in the \textsf{Enter Data} tab must be integers or finite
real floats. A fraction, a floating point infinity, a nan or a non-real
number is not accepted. If an imaginary part of your data is negligible
noise, then replace it with 0.0. Otherwise you must separately fit
the real and imaginary parts or the absolute value and the arg, which
is the \textsf{Two-Argument Arctangent} in Eureqa. If your data has
an undefined value due to an indeterminate form, then replace it with
the value returned by your computer algebra $\mathrm{limit}\,(\ldots)$
function. If your table has an infinity, then subtract out or divide
out the singularity that is causing it. Don't expect a good fit if
you simply omit a value that has infinite magnitude or replace it
with an exceptionally large magnitude having the correct sign -- either
of which can dramatically mislead the search.\vspace{-3pt}

\item I entered $1-t^{2}$ as $(1-t)(1+t)$ in input (\ref{eq:xyPairs})
and elsewhere to reduce the magnification of rounding errors by catastrophic
cancellation near $x=\pm1$.\vspace{-3pt}

\item The reason for $\mathrm{WorkingPrecision}\rightarrow25$ is that I
wanted to further reduce this catastrophic cancellation.\vspace{-3pt}

\item A faster but less accurate way to compute a table for a numeric antiderivative
is to use a numeric ODE solver, but the time consumed by $\mathrm{NIntegrate}[\ldots]$
was negligible compared to the time consumed by Eureqa and the much
greater time consumed by me.
\end{itemize}
}
\item All the values were real and finite, so I then entered\vspace{-5pt}
\[
\mathrm{Export}\left[\mathrm{"antiderivative.csv"},\,\mathrm{xyPairs}\right],\]
then I imported the resulting file into Eureqa and tried to fit $W=f(x)$
with the default formula building blocks. After three minutes of search
the most accurate formula was\vspace{-5pt}
\begin{eqnarray*}
W & = & \dfrac{0.7773112536\, x}{\cos(0.4696613973\, x^{7})-0.2294845216\, x^{2}},\end{eqnarray*}
which is non-insightful and accurate to a disappointing maximum absolute
error of about 0.4\%.\vspace{-5pt}

\item A \textsl{Mathematica} plot of the integrand reveals a probable explanation:
The integrand has endpoint singularities, and although they are integrable,
the resulting integral has infinite magnitude low-order derivatives
at the endpoints, which probably makes the quadrature less accurate
and the Eureqa search slower than otherwise.\vspace{-5pt}

\item The plot of numeric antiderivative values on the \textsf{Prepare Data}
tab revealed that they look approximately proportional to $\mathrm{\, asin}(x)$.\vspace{-5pt}

\item So next I did another Eureqa search after disabling the sine and cosine
building blocks and making the target expression be $W=f(x\mathrm{,asin}(x))$.
I did \textsl{not} enable the arcsine building block because, for
example, I didn't want Eureqa to waste time trying sub-expressions
such as $\mathrm{asin}(1+3x^{2})$. In 11 seconds Eureqa found the
following seven term expression having a maximum absolute error of
only about $3.1\times10^{-9}$:\vspace{-5pt}
\begin{multline}
W\:=\:-1.870027576\times10^{-13}+0.7816747744\, x+0.0147770774\, x^{3}-\\
0.03033616234\, x^{2}\mathrm{asin}(x)+0.07586494202\, x\mathrm{\, asin}(x)^{2}+\\
0.0818165982\,\mathrm{asin}(x)^{3}+0.0009144579166\, x^{3}\mathrm{asin}(x)^{2}.\quad\label{eq:MorePrecise}\end{multline}
I aborted the search at 3 minutes with no further improvement. Regarding
this result:\vspace{-5pt}

\begin{enumerate}
\item To view the coefficients rounded to 10 significant digits rather than
about 4, I had to copy the expression from the \textsf{Best solutions
of different sizes} pane in the \textsf{View Results }tab into \textsl{Mathematica}
or a text editor. Unfortunately there is no way to view all 16 digits,
so I will have to polish a result with other software to obtain a
silver standard result.%
\footnote{The rounding of coefficients in the \textsf{Project Log} and \textsf{Best
solutions of different sizes} panes \textsl{does} increase comprehensibility
-- particularly since those panes are regrettably un-scrollable. Even
more rounding could justifiably be done for coefficients of terms
whose relative maximum contribution is small. For example, if the
maximum contribution of a term over all data points is 1\%, then its
coefficient justifiably could be rounded to 2 less significant digits
than the coefficient that contributes the most over all data points.
See \cite{RoundingCoefficients} for more about this idea.%
}\vspace{-3pt}

\item I edited the copied formula because it contained annoying superfluous
parentheses and represented integer powers by repeated multiplication.\vspace{-3pt}

\item The even symmetry of the integrand and the centered integration from
0 imply that the antiderivative should have odd symmetry. Therefore
the spurious constant term in expression (\ref{eq:MorePrecise}) should
be discarded. Terms that are contrary to odd symmetry might arise
from the random partitioning of the data into training and validation
sets.\vspace{-3pt}

\item Converting the result to recursive form by collecting similar powers
of $x$ or $\mathrm{asin}(x)$ can significantly reduce the \textsf{Complexity}.
For example, rounding coefficients the coefficients in result (\ref{eq:MorePrecise})
for brevity,\begin{multline*}
W\;=\;0.78\, x+0.015\, x^{3}-0.03\, x^{2}\mathrm{asin}(x)+\\
(0.076\, x+0.00091\, x^{3})\mathrm{\, asin}(x)^{2}+0.082\mathrm{\, asin}(x)^{3}.\end{multline*}
which saves one instance of {}``$*\mathrm{asin}(x)^{2}$. At the
expense of legibility, the \textsf{Complexity} and the floating-point
substitution time can be further reduced by \textsl{Hornerizing} this
recursive representation to\vspace{-5pt}
\begin{multline*}
W\;=\; x(0.78+0.015x^{2})+\\
\mathrm{asin}(x)(-0.03\, x^{2}+(x\,(0.076+0.00091\, x^{2})+0.082\,\mathrm{asin}(x))\mathrm{\, asin}(x))).\end{multline*}
This is another place where more computer algebra could help Eureqa.\vspace{-3pt}

\item Notice that result (\ref{eq:MorePrecise}) has no term that is simply
a numeric multiple of $\mathrm{asin}(x)$. Therefore it doesn't model
the infinite endpoint slopes associated with the integrand singularities
there. But that is my fault because the target expression contains
no special encouragement to include a term of that form, and the plot
from the Eureqa \textsf{Prepare Data} tab reveals that the sample
points were not closely enough spaced near the endpoints to suggest
the infinite slope magnitudes there.\vspace{-5pt}

\end{enumerate}
\item Result (\ref{eq:MorePrecise}) suggests that a set of particularly
good fits would be truncated series of the form\vspace{-5pt}
\[
\left(c_{1}\,\mathrm{asin}\,(x)+c_{2}\, x\right)+\left(c_{3}\,\mathrm{asin}\,(x)^{3}+c_{4}\, x\mathrm{\, asin}\,(x)^{2}+c_{5}\, x^{2}\mathrm{asin}\,(x)+c_{6}\, x^{3}\right)+\cdots,\]
which can be regarded as a truncated bi-variate series in terms having
a power of $x$ times a power of $\arcsin(x)$ that have non-decreasing
odd total degrees, with numeric coefficients $c_{k}$. I could constrain
Eureqa to search only within this class of expressions by entering
a target expression having a particular total degree, such as\vspace{-5pt}
\[
W\!=\! f_{1}()\,\mathrm{asin}\,(x)+f_{2}()\, x+f_{3}()\mathrm{\, asin}\,(x)^{3}+f_{4}()\, x\mathrm{\, asin}\,(x)^{2}+f_{5}()\, x^{2}\mathrm{asin}\,(x)+f_{6}()\, x^{3}.\]
However, standard regression software is much faster for merely optimizing
some parameters in a specific form -- and more accurate if done with
arbitrary-precision arithmetic.%
\footnote{This is another area where other kinds of mathematical software could
help Eureqa. Eureqa would probably be much faster if it used standard
linear or nonlinear regression while it is merely \textsl{adjusting}
coefficients in a particular form. Now that Eureqa has demonstrated
how well genetic search can do unassisted, there is no reason not
to make it much faster with assistance from other disciplines.%
} The largest total degree in Eureqa result (\ref{eq:MorePrecise})
is 5. Consequently I next used the \textsl{Mathematica} $\mathrm{LinearModelFit}\,[\ldots]$
function with the 12 basis expressions of the form $x^{j}\mathrm{asin}\,(x)^{k}$
having odd total degree $1\leq j+k\leq5$. After 0.27 seconds I received
the following eight term result whose significant digits I have truncated
for brevity:\vspace{-5pt}
\begin{multline}
22.9\, x+0.012\, x^{3}+0.000097\, x^{5}-22.1\mathrm{\, asin}\,(x)-0.068\, x^{2}\mathrm{asin}\,(x)+\\
0.78\, x\mathrm{\, asin}\,(x)^{2}+3.09\mathrm{\, asin}\,(x)^{3}-0.078\mathrm{\, asin}\,(x)^{5}.\quad\label{eq:LinearModelFit5}\end{multline}
The maximum absolute error at the tabulated points was about $4\times10^{-11}$,
which is two more significant digits than (\ref{eq:MorePrecise})
for the same total degree and only one more term. Very gratifying.\vspace{-5pt}

\item Encouraged, I next tried $\mathrm{LinearModelFit}\,[\ldots]$ with
total degree 7, which entails 20 basis terms. After 0.52 seconds I
received a different eight term result\vspace{-5pt}
\begin{multline}
7.48\, x+0.0041\, x^{3}-6.7\,\mathrm{asin}\,(x)-0.019\, x^{2}\mathrm{asin}\,(x)+0.26\, x\mathrm{asin}\,(x)^{2}-\\
0.000012\, x^{5}\mathrm{asin}\,(x)^{2}+1.01\,\mathrm{asin}\,(x)^{3}-0.025\,\mathrm{asin}\,(x)^{5}\quad\label{eq:LinearModelFit7}\end{multline}
having an absolute error at the tabulated points of about $1.5\times10^{-11}$.
Thus it seems likely that something important is missing from the
basis, preventing results (\ref{eq:LinearModelFit5}) and (\ref{eq:LinearModelFit7})
from corresponding to economized truncations of a convergent infinite
series because:\vspace{-5pt}

\begin{enumerate}
\item This is a negligible accuracy improvement.\vspace{-3pt}

\item The term count didn't increase.\vspace{-3pt}

\item The coefficients of similar terms change dramatically between approximations
(\ref{eq:LinearModelFit5}) and (\ref{eq:LinearModelFit7}).\vspace{-3pt}

\item In each approximation there is substantial cancelation between the
dominant terms.\vspace{-5pt}

\end{enumerate}
\end{enumerate}
Thus it is not promising to try larger total degrees with this basis
in pursuit of a concise numerically stable silver standard. Nonetheless,
this has been a nice combined use of Eureqa and \textsl{Mathematica}:
After the decision to try a target expression of the form $f(x,\mathrm{asin}(x))$,
Eureqa discovered a concise form having 9 significant digits and basis
functions of the form $x^{j}\mathrm{asin}(x)$. I then used \textsl{Mathematica}
to obtain a result having only one more term that is accurate to 11
significant digits, which is more than adequate for many purposes.
For this example Eureqa has served as a muse rather than an oracle.

However, I couldn't help wondering: \textsl{Is} there a relatively
simple class of forms that Eureqa might have revealed for further
polishing to a concise \textsl{silver-standard} fit by\textsl{ Mathematica}?

With the help of the \textsl{Mathematica} \textsl{$\mathrm{Series}\,[\ldots]$
}function and an embarrassingly large amount of my time adapting its
results to my desires via a sixteen line \textsl{Mathematica} function
that I wrote, I was able to determine an exactly integrable truncated
infinite series expansion of the integrand that converged sufficiently
fast without undue catastrophic cancelation over the entire interval
$-1\leq x\leq1$. Using $\mathrm{Assuming}[-1\leq x\leq1,\mathrm{Integrate}[\ldots]]$,
the Hornerized representation of the corresponding \textsl{antiderivative}
is\vspace{-5pt}
\begin{eqnarray}
\tilde{W}(x) & := & c_{0}\,\mathrm{asin}(x)+x\sqrt{(1-x)(1+x)}\left(c_{1}+x^{2}\left(c_{2}+x^{2}(c_{3}+\cdots)\right)\right).\label{eq:BifocalSeries}\end{eqnarray}
Being a single series bi-focally expanded about $x=\pm1$ rather than
a conditional expression containing two or more different series,
it is $C^{\infty}$ over the entire interior interval $-1<x<1$. Notice
that contrary to the previous efforts, $\mathrm{asin}(x)$ occurs
only to the first power and that all the other terms are multiplied
by $\sqrt{(1-x)(1+x)}$. A good set of basis expressions is thus $\{\mathrm{asin}(x),\, x^{1+2k}\sqrt{(1-x)(1+x)}\}$
for $k=0,1,\ldots$. The higher powers of $\mathrm{asin}\,(x)$ in
the previous basis were being used as flawed surrogates for $x^{1+2k}\sqrt{(1-x)(1+x)}$:
Octagonal pegs in circumscribed round holes -- they sort of fit, but
in an impaired way .

The \textsl{Mathematica} function that I wrote computes the exact
coefficients $c_{k}$, which involve $\sqrt{6}$ and powers of $\pi$.
However, I instead used $\mathrm{LinearModelFit}\,[\ldots]$ to effectively
economize a higher degree approximation to a lower-degree one having
nearly the same accuracy. For the data I used high precision numerical
integration after using truncated series (\ref{eq:BifocalSeries})
to subtract out and exactly integrate two terms of the endpoint singularities.
This made equally spaced $x$ values quite acceptable. Experiments
revealed that using terms through coefficient $c_{7}$ gave an approximation
to $y(x)$ that had a worst absolute error of about $6\times10^{-16}$
at the right endpoint where the value was about 1.255. A plot of the
residuals revealed that this eight-coefficient approximation has a
relative error of about $5\times10^{-16}$, making it silver standard.
Moreover, this plot was essentially noise, revealing no remaining
exploitable residual for IEEE double.

Eureqa did not by itself discover this better basis, but that is my
fault: I never allowed square roots as a Building block. I should
have done a little analysis earlier to wisely suggest a target expression
of the form $f(x,\mathrm{\, asin}(x),\,\sqrt{(1-x)(1+x)})$. After
all, there is a square root in the integrand denominator, so I should
have expected a square root in a good approximate antiderivative numerator.
I could have even tried the square root that also contains the cosine
term. Using Eureqa effectively for computer algebra and numerical
analysis should be viewed as a collaboration rather than a competition.
And Eureqa is not a black box oracle. You cannot leave your brain
behind. That is probably equally important when using Eureqa for noisy
experimental data.

\subsection{Other possible approximate symbolic results}

There are many other possibilities for using Eureqa to find a concise
expression or suggest a good class of expressions that closely approximates
a result that would otherwise have to be presented only graphically
or as a table. For example,\vspace{-5pt}

\begin{itemize}
\item \textsl{Inverse functions and solving parametric algebraic equations}:
Imagine that for the example of the previous subsection, we also or
instead wanted an approximate symbolic expression for the \textsl{inverse}
Dogbert $W$ function: $x$ as a function of $W$. Eureqa can search
for such expressions if we simply enter $x=f(W)$, or perhaps more
specifically for this example $x=\sin(f_{2}()\, W)+f_{3}(W)$.\vspace{-5pt}

\item \textsl{Solution of differential, integral, delay and other kinds
of functional equations}: For example, suppose that we have a system
of first-order ordinary differential equations, and a tabulated numerical
solution vector $[y_{1}(t),y_{2}(t),\ldots]$ for $t=t_{0},t_{1},\ldots t_{n}$.
Then we can use Eureqa to separately search for good expressions for
$y_{1}(t),y_{2}(t),\ldots$. If we want $t_{n}=\infty$, then we should
first make a change of independent variable to map that end point
to a finite value in a way that doesn't introduce a singularity elsewhere.\vspace{-5pt}

\item \textsl{Implicitization}: A major application of computer algebra
is implicitization of parametric equations defining curves, surfaces
or higher-dimensional manifolds. However, exact methods are known
only for certain classes of parametric equations. For other classes
we can try using Eureqa to find approximate implicit equations. (Hint:
see \cite{SchmidtAndLipsonImplicitEqn} for tips on how to use Eureqa
for this purpose.)
\end{itemize}

\section{Additional tips and tricks}

Here are a few additional tips and tricks for using Eureqa effectively
in conjunction with computer algebra or numerical analysis:\vspace{-5pt}

\begin{enumerate}
\item Building blocks such as $\mathrm{asin}(\ldots)$, $\mathrm{acos}(\ldots)$,
$\mathrm{atanh}(\ldots)$, $\log(\ldots)$, and $\mathrm{pow}(\ldots,\ldots)$
that are not real for some real arguments cause a candidate to be
rejected if the result isn't real for \textsl{all} of the data values.
To avoid wasting time partially processing such candidates, consider
using a specific form of such a function in your target expression
that automatically precludes unreal sub-expressions rather than simply
including the function as a building block. For example, knowing that
$-1\leq x\leq1$, I used $W=f(x,\mathrm{asin}(x))$ for the example
in subsection 3.1 rather than merely checking the Arcsine building
block. \vspace{-5pt}

\item Conversely, your target expression can be of the form {}``$\ldots=f(x)+0*\log(f(x)-7)$''
to impose an inequality constraint of the form $f(x)>7$. Whenever
$f(x)\leq7$ the log factor generates a nan, and 0{*}nan is nan, which
disqualifies that candidate. Otherwise the multiplication by 0 makes
that term have no effect on the search for good $f(x)$, and we can
mentally discard the artifice log term at the end. Similarly a target
expression of the form {}``$\ldots=f(x)+0*\mathrm{asin}\left((f(x)-7)/3\right)$''
imposes the inequality constraint $4\leq f(x)\leq10$. (Perhaps a
future version will have a more efficient and straightforward way
to impose inequality constraints.)\vspace{-5pt}

\item To increase the chance of obtaining exact rational coefficients, reduce
your input expression over a common denominator, then separately tabulate
and fit the numerator and denominator with the \textsf{Integer Constant}
building block checked.\vspace{-5pt}

\item To possibly include \textsl{foreseen} exact symbolic known irrational
constants such as $\pi$ and $\sqrt{2}$ in your result, for each
such constant make a labeled column of corresponding floating-point
values, preferably accurate to 16 digits, For example if a column
labeled $\mathit{pi}$ contains $3.14159\ldots$ and a column labeled
$\mathit{sqrt}2$ contains $1.41421\ldots$, then check the \textsf{Integer
Constant} building block and use a target expression of the form {}``$\ldots=f(pi,\mathit{sqrt}2,\ldots)$''.
These symbolic constants don't contribute much to the curse of dimensionality
because they are invariant, so there is no need to include more rows
on their account.\vspace{-5pt}

\item As with most regression software, Eureqa tends to express results
in terms of the notoriously ill-conditioned monomial basis $1,x,x^{2},x^{3},\ldots$
To express your result more accurately in terms of an orthogonal polynomial
basis, such as the Chebychev $T$ polynomials, generate columns of
the desired number of successive polynomials at the tabulated values
of the independent variables labeled $T_{0},T_{1},\ldots$, then use
a target expression of the form {}``$\ldots=f(T_{0},T_{1},\ldots)$''.
These numerous $T_{k}$ variables don't contribute much to the curse
of dimensionality because they are totally correlated rather than
truly independent. Moreover, these variables have the same \textsf{Complexity}
as any other variable, thus preventing them from being discriminated
against as they would be if you otherwise equivalently made the target
expression be {}``$\ldots=f(1,x,x^{2}-1,x^{3}-3x,\ldots)$''.\vspace{-5pt}

\item Eureqa has a built-in function $D(\mathit{columnVariable}_{1},\mathit{columnVariable}_{2},n)$
that uses numerical differentiation of a smoothing spline to approximate
an $n$th derivative. One of its most common uses is for system identification:
Given experimental tabulated $t$ $y$ pairs, if you enter a target
expression such as $D(y,t,2)=f(y,D(y,t,1),t)$, then Eureqa will search
for a concise differential equation of that rather general form that
accurately fits the data. For example, the returned results might
include $D(y,t,2)=x^{2}y^{2}D(y,t,1)+y^{3}+\cos(t)$. Although differentiation
of a smoothing spline is probably significantly more accurate than
more naive methods -- particularly for noisy data -- numerical differentiation
inherently magnifies rounding and discretization errors. Therefore
if an explicit or implicit expression is available for $y$, then
it is almost certainly more accurate to tabulate floating-point values
of exact derivatives produced by computer algebra. Automatic differentiation
\cite{automaticDifferentiation} is another silver-standard alternative.
\vspace{-5pt}
 
\item Eureqa has a built-in $\mathrm{integral}(\ldots)$ function that I
could have used it instead of the \textsl{Mathematica} $\mathrm{NIntegrate}[\ldots]$
function. However, the $\mathrm{integral}(\ldots)$ function uses
the trapezoidal rule -- probably because it has no control over the
spacing of the independent variable values. Given an integrand expression
to integrate exactly or to sample adaptively as needed by a function
such as $\mathrm{NIntegrate}(\ldots)$, computer algebra can almost
always deliver a more accurate integral.\vspace{-5pt}

\end{enumerate}

\section{The need for seamless integration of separate software packages}

It is usually somewhat of a nuisance to work between two or more separate
software packages compared to working entirely within one that has
all of the necessary features built in. The extra effort and the lesser
chance of even knowing about independent software packages is a great
deterrent to such use by the masses. Thus it is good to know that
interfaces are under development for using Eureqa from within \textsl{Mathematica},
Matlab, Python, .NET and KNIME, with current versions down-loadable
from \cite{Schmidt}. Such added interfaces are rarely as seamless
as features that are built in, but they are often great improvements
over communicating via export-import of files or copy and paste. Among
other things, the interfaces can accommodate differences in syntax
for expressions.

\section{The need for automation}

At any one time, most of us are amateurs with most of the software
that we use. We need all of the help that we can obtain. In response
to that need, many of the best software packages are automating portions
that can be. For example:\vspace{-5pt}

\begin{enumerate}
\item Adaptive quadrature relieves us of having to learn about numerous
rules for different kinds of univariate and multivariate integrands
and finite, semi-infinite or infinite integration regions.\vspace{-5pt}

\item Adaptive interval and significance arithmetic can give us results
that are guaranteed or highly likely to have the accuracy that we
request.\vspace{-5pt}

\item Eureqa can sometimes automatically discover good regression models.\vspace{-5pt}

\item Inverse Symbolic Calculator can sometimes automatically recognize
floating-point approximations to exact symbolic rational or irrational
constants.\vspace{-5pt}

\end{enumerate}
However, the examples in subsections \ref{sub:A-trigonometric-simplificationEg}
and \ref{sub:An-antiderivative-example} were \textsl{not} automatic.
Training, experience and judgment were involved in choosing the data
points, target expression, etc. Some of this could and should be automated.
A qualitative analysis program such as the Maxima program described
in \cite{QualitativeAnalysis} could help in these regards: Given
an expression, this program attempts to return bounds and indications
of monotonicity, convexity, symmetries, periodicities, zeros, singularities,
limits and stationary points. To the extent that this is successful,
this information could be used to choose automatically for each independent
variable the endpoints, the number of samples, and perhaps even their
distribution. For example:\vspace{-5pt}

\begin{itemize}
\item If there are singularities, then it is best to automatically convert
any tangents to sines divided by cosines, then form a reduced common
denominator, then use Eureqa to fit separately the numerator and denominator.
Neither the numerator nor denominator will contain a pole, but they
still could contain a logarithmic singularity that would have to be
handled by subtracting or dividing it out. \vspace{-5pt}

\item It is probably best if the endpoints extend modestly beyond all of
the real zeros and stationary points, including at least one fundamental
period if any, with enough points to resolve the shortest period.\vspace{-5pt}

\item If there is a symmetry and the expression is $C^{\infty}$ at the
symmetric point, then it is probably best to center the data values
on that point. Otherwise it might be more efficient to make that symmetry
point be one of the end points.\vspace{-5pt}

\end{itemize}
Even a purely \textsl{numeric} program that searched for zeros, singularities,
extrema, periodicities, and symmetries could help in these regards.\vspace{-5pt}

\section{Conclusions}
\begin{enumerate}
\item With wise use, Eureqa can sometimes determine a simpler exact equivalent
expression than current computer algebra systems can -- or sometimes
transform an expression into a targeted form that isn't provided by
a computer algebra system.\vspace{-5pt}

\item With wise use, Eureqa can sometimes suggest promising forms of expressions
that approximate a result for which an exact closed form is unobtainable
or excessively complicated. Often you will want to obtain a more accurate
result in a thus-revealed class by using a linear or non-linear regression
program built into a computer algebra system or statistics package.\vspace{-5pt}

\item Some simple interface additions \textsl{within} Eureqa could increase
its utility for the above two purposes. Interfaces \textsl{to} Eureqa
within more other software packages would encourage more use of Eureqa.
Embedding Eureqa within those packages would probably be even more
seamless.\vspace{-5pt}

\item Eureqa uses some custom computer algebra and perhaps some standard
numerical methods internally. Eureqa would almost certainly benefit
from embedding and exploiting a well developed full-fledged computer
algebra system together with classic regression and numerical methods
packages.\vspace{-5pt}

\item Eureqa's most frequent and notable successes will probably continue
to be with noisy experimental data, but Eureqa shows promise for purposes
1 and 2 above.\vspace{-5pt}

\end{enumerate}

\section*{Acknowledgment}

Thank you Michael Schmidt for your suggestions and patient explanations.

\section*{Figures}

\noindent \begin{center}
\begin{figure}[H]
\caption{Eureqa \textsf{Start Search }tab for exact trigonometric simplification
example}

\includegraphics[bb=0bp 0bp 679bp 695bp]{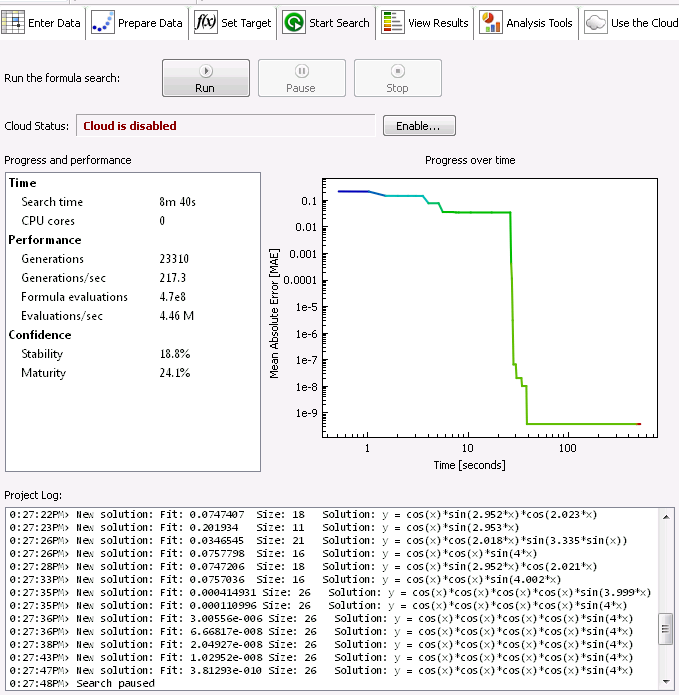}\label{Flo:StartSearchTab}
\end{figure}

\par\end{center}

\begin{figure}[H]
\noindent \centering{}\caption{Eureqa \textsf{View Results }tab for exact trigonometric simplification
example}
\label{Flo:ViewResultsTab}\includegraphics[bb=0bp 0bp 536bp 540bp]{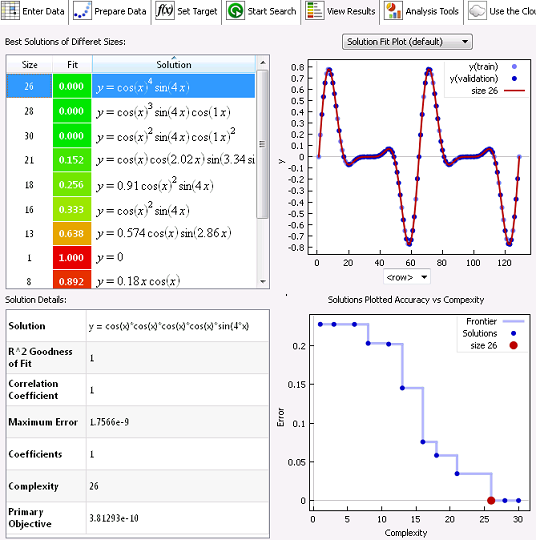}
\end{figure}

\end{document}